\documentclass[aps,preprint,showpacs,nofootinbib]{revtex4}
\usepackage{amsfonts}
\usepackage{amsmath}
\usepackage{amssymb}
\usepackage{relsize}
\usepackage{graphicx}
\usepackage{epsfig}
\usepackage{float}
\usepackage[caption = false]{subfig}
\usepackage{color}

\usepackage{multirow}
\begin{document}

\title{Analytic matrix elements for the two-electron atomic basis with logarithmic terms}

\author{Evgeny Z. Liverts}
\affiliation{Racah Institute of Physics, The Hebrew University, Jerusalem 91904,
Israel}

\author{Nir Barnea}
\affiliation{Racah Institute of Physics, The Hebrew University, Jerusalem 91904,
Israel}

\begin{abstract}
The two-electron problem for the helium-like atom/ions in $S$-state is considered.
The basis containing the integer powers of $\ln r$, where $r$ is a radial variable of the Fock expansion, is studied.
In this basis, the analytic expressions for the matrix elements of the corresponding Hamiltonian are presented.
These expressions include only special functions presented by the built-in \emph{Mathematica} codes,
what enables very fast and accurate computation of the matrix elements.
The decisive contribution of the correct logarithmic terms to the behavior of the two-electron wave
function in the vicinity of the triple-coalescence point is reaffirmed.
\end{abstract}


\maketitle

\section{Introduction}\label{S0}
The two-electron atom/ions present an opportune system both for
studying a number of photoelectron and other atomic processes, and for
testing the new quantum calculational approaches.
The proper variational calculations using conventional expansions containing exponential and polynomial functions of the interparticle distances $r_1, r_2$ and $r_{12}$ have been studied since the classic work of Hylleraas \cite{HYL}.
However, a few years later, Bartlett et al \cite{B35} have shown that these analytic expansions of the wave function $\Psi$ cannot satisfy the Schr\"{o}dinger equation for the helium-like system.
Bartlett \cite{B37} then showed that the expansion about origin can be formally presented as
$\Psi=\sum_{k=0}^\infty C_k(r_1,r_2,r_{12})(\ln r)^k$, with $r=\sqrt{r_1^2+r_2^2}$.
Fock \cite{FOCK} proposed that, in the neighbourhood of $r=0$, the exact eigenfunctions have the (Fock) expansion
\begin{equation}\label{1}
\Psi(r,\alpha,\theta)=\sum_{k=0}^{\infty}r^k\sum_{p=0}^{[k/2]}\psi_{k p}(\alpha,\theta)(\ln r)^p,
\end{equation}
where
$ \alpha=2\arctan \left(r_2/r_1\right)$ and $\theta=\arccos\left[(r_1^2+r_2^2-r_{12}^2)/(2r_1r_2)\right]$ are the hyperspherical angles.

The electron structure studies incorporating $\ln (r_1+r_2)$ that is closely related to $\ln r$, have been then presented by Hylleraas abd Midtdal \cite{HYM}, Frankowski and Pekeris \cite{FRP}, and Freund et al \cite{FHM}.
The exact form of the Fock expansion truncated to 32 (and/or 52) terms, have been studied by Ermolaev and Sochilin \cite{ES1}.
Abbott, Gottschalk and Maslen \cite{GAM} have presented a comprehensive analysis of techniques for solving the Fock expansion.
Compact representation of helium wave function, that includes an explicit treatment of the Fock expansion supplemented with the Laguerre functions basis, has been discussed by Forrey \cite{FOR} (see also references therein).

We have shortly presented the basic works that treated the Fock logarithmic terms for the electron structure calculations of the two-electron atoms.
Only a few of them (in fact \cite{ES1} and \cite{FOR}) used the basis containing $\ln r$,
what is the necessary condition (according to the Fock expansion (\ref{1})) for obtaining the two-electron wave function with correct behavior near the point $r=0$.
The latter presents, so called, triple coalescence point playing an important role in many physical processes treated by the atomic and related fields of physics (see, e.g., \cite{MYE,DRU}).

The vast majority of authors (see, \cite{DRK} and references therein) avoid to use basis containing $\ln r$, or
at best they substituted it by $\ln (r_1+r_2)$.
The reason is the calculational/computational difficulties in the absence of a significant gain in accuracy of the resulting energy values.

Given the mentioned above importance of accurate calculation of the two-electron wave function in the vicinity of the triple-coalescence point, we present here the analytic expressions for the matrix elements of the proper Hamiltonian, for the case of the basis containing $r,~\ln r$ and their integer powers.

\section{Matrix Elements of the Hamiltonian}\label{S1}
We will operate with the coordinates system, first introduced by Hylleraas:
\begin{equation}\label{2}
s=\delta(r_1+r_2),~~~t=\delta(r_1-r_2),~~~u=\delta r_{12},
\end{equation}
where $\delta$ is variational parameter. The volume element for the coordinates system (\ref{2}) is:
\begin{equation}\label{3}
d V=\delta^{-6}\pi^2\left(t^2-s^2\right)u.
\end{equation}
The atomic units is used throughout the paper.
The kinetic energy operator $-\nabla^2/2$ for the $S$-state two-electron problem with infinite nucleus mass has the following form in the $\{s,t,u\}$ coordinates (see, e.g., \cite{GAM}):
\begin{eqnarray}\label{4}
\frac{\nabla^2}{2\delta^2}=\frac{\partial^2}{\partial s^2}+\frac{\partial^2}{\partial t^2}+\frac{1}{u^2}\frac{\partial}{\partial u}u^2\frac{\partial}{\partial u}+\frac{4}{s^2-t^2}\left(s\frac{\partial}{\partial s}-t\frac{\partial}{\partial t}\right)+
~\nonumber~~~~~~~~~~~~~~~~~~~~~~~~~~~~~~~~~~~~~\\
\frac{2}{u(s^2-t^2)}\left[s(u^2-t^2)\frac{\partial^2}{\partial s \partial u}-t(u^2-s^2)\frac{\partial^2}{\partial t \partial u}\right].~~~~~
\end{eqnarray}
The Hamiltonian $\mathcal{H}=-\nabla^2/2+U$ includes
the Coulomb potential
\begin{equation}\label{5}
U\equiv -\frac{Z}{r_1}-\frac{Z}{r_2}+\frac{1}{r_{12}}=\delta\left(\frac{4 Z s}{t^2-s^2}+\frac{1}{u}\right),~~~~~~~~~~~~~~~~~~~~~~~~~~~~~~~~~~~
\end{equation}
where $Z$ is a nucleus charge.

In accordance with what was stated in Introduction, for singlet $S$-states we select the basis functions of the form:
\begin{equation}\label{6}
\phi_{n,l,m,i,j}(s,t,u)=\exp\left(-\frac{s}{2}\right)s^n t^{2l} u^m (s^2+t^2)^{i/2} \ln^j r,
\end{equation}
where $n,l,m,i,j$ are integers, and  $r=\sqrt{(s^2+t^2)/2}$.
Index $i$ can take the values $-1,0,1$.
This basis is very close to the Frankowski basis \cite{FRP}, but with replacing $\ln s$ by $\ln r$,
and considering the additional values $i=-1$ (with $j=0$), that is needed to take into consideration (at least) the term $u^3/r$ of the Fock expansion (see, e.g., \cite{FRP}).

Our aim is to solve the Schr\"{o}dinger equation
\begin{equation}\label{7a}
\left(\mathcal{H}-E\right)\Psi=0,
\end{equation}
where
\begin{equation}\label{7}
\Psi(s,t,u)=\sum_{n,l,m,i,j}C(n,l,m,i,j)\phi_{n,l,m,i,j}(s,t,u)
\end{equation}
is the two-electron atomic wave function. The approximate value of the total energy $E$, as well as the expansion coefficients, $C(n,l,m,i,j)$ can be found as solution of the generalized eigenvalue equation
\begin{equation}\label{8}
(\delta^2\textbf{K}+\delta\textbf{U})\textbf{C}=E \textbf{S} \textbf{C},
\end{equation}
where $\textbf{K}$ and $\textbf{U}$ are the matrix representations of the operators $-\nabla^2/2\delta^2$ and $U/\delta$, respectively, for the basis (\ref{6}) and the coordinates system (\ref{2}).
Here $\textbf{C}$ presents eigenvector corresponding to the eigenvalue $E$ of the matrix equatian (\ref{8}), whereas $\textbf{S}$ is the overlap matrix for the basis (\ref{6}).

Using Eq.(\ref{3}), one obtains for the entries of the overlap matrix:
\begin{eqnarray}\label{9}
S(n',l',m',i',j'|n,l,m,i,j)=\widetilde{S}(N,L,M,I,J)=
~\nonumber~~~~~~~~~~~~~~~~~~~~~~~~~~~~~~~~~~~~~\\
\int_0^\infty ds\int_0^s du\int_0^u dt e^{-s}s^{N}t^{2L}u^{M+1}(t^2-s^2)\left(s^2+t^2\right)^{I/2}
\left( \ln \frac{s^2+t^2}{2}\right)^{J}=
\nonumber~~~\\
\mathcal{P}_{I,J}(N,2L+2,M+1)-
\mathcal{P}_{I,J}(N+2,2L,M+1),~~
\end{eqnarray}
where
\begin{equation}\label{10}
N=n'+n,~L=l'+l,~M=m'+m,~I=i'+i,~J=j'+j.
\end{equation}
We introduced the following notation:
\begin{equation}\label{11}
\mathcal{P}_{\imath,\jmath}(\nu,\ell,\mu)=\int_0^\infty ds\int_0^s du\int_0^u dt e^{-s}s^{\nu}t^{\ell}u^{\mu}\left(s^2+t^2\right)^{\imath/2}
\left( \ln \frac{s^2+t^2}{2}\right)^{\jmath}.
\end{equation}
Using Eq.(\ref{5}) for the Coulomb potential, one can express the entries of the $\textbf{U}$-matrix in terms of the integrals (\ref{11}), as follows:
\begin{eqnarray}\label{12}
\widetilde{U}(N,L,M,I,J)=
~\nonumber~~~~~~~~~~~~~~~~~~~~~~~~~~~~~~~~~~~~~~~~~~~~~~~~~~\\
\int_0^\infty ds\int_0^s du\int_0^u dt e^{-s}s^{N}t^{2L}u^{M+1}\left(\frac{4Zs}{t^2-s^2}+
\frac{1}{u}\right)(t^2-s^2)\left(s^2+t^2\right)^{I/2}
\left( \ln \frac{s^2+t^2}{2}\right)^{J}=
\nonumber~~~\\
4Z\mathcal{P}_{I,J}(N+1,2L,M+1)+\mathcal{P}_{I,J}(N,2L+2,M)-\mathcal{P}_{I,J}(N+2,2L,M).~~~~~~~~~~~~~~
\end{eqnarray}
We will consider the basis functions (\ref{6}) with the maximum power of $\ln r$ equals $2$, that is $j=0,1,2$. In accordance with the latter,
in order to express the entries of the $\textbf{K}$-matrix in terms of the integrals (\ref{11}), it is convenient to split this entries into groups
\begin{equation}\label{13}
 K_{i,j}\equiv\int_0^\infty ds\int_0^s du\int_0^u dt~\phi_{n',l',m',i',j'}(s,t,u)u(t^2-s^2)\frac{\nabla^2}{2\delta^2}\phi_{n,l,m,i,j}(s,t,u).
\end{equation}
We restrict ourselves to the cases $i=0,1$; $j=0,1,2$, and the additional case $i=-1,j=0$.
 Then, introducing notation
\begin{equation}\label{13a}
P_{\Delta_{i'},\Delta_{j'}}(\alpha,\beta,\gamma)=\mathcal{P}_{i'+\Delta_{i'},j'+\Delta_{j'}}\left(N+\alpha,2L+\beta,M+\gamma\right),
\end{equation}
 and using Eq.(\ref{4}) and definition (\ref{11}) one obtains:
\begin{equation}\label{14}
K_{0,0}=\sum_{k=1}^{10}b_k^{0,0}(n,l,m)P_{0,0}\left(\alpha_k^{0,0},\beta_k^{0,0},\gamma_k^{0,0}\right)
,~~~~~~~~~~~~~~~~~~~~~~~~~~~~~~~~~~~~~~~~~~~~~~~~~
\end{equation}

\begin{eqnarray}\label{15}
K_{0,1}=\sum_{k=1}^{10}b_k^{0,0}(n,l,m)P_{0,1}\left(\alpha_k^{0,0},\beta_k^{0,0},\gamma_k^{0,0}\right)+
2\left[P_{-2,0}\left(3,0,1\right)-P_{-2,0}\left(1,2,1\right)\right]+
~\nonumber~~~~\\
4(2l+m+n+2)\left[P_{-2,0}\left(0,2,1\right)-P_{-2,0}\left(2,0,1\right)\right],~~~~~~~~~~~~~~
\end{eqnarray}

\begin{equation}\label{16}
K_{1,0}=\sum_{k=1}^{14}b_k^{1,0}(n,l,m)P_{-1,0}\left(\alpha_k^{1,0},\beta_k^{1,0},\gamma_k^{1,0}\right)
,~~~~~~~~~~~~~~~~~~~~~~~~~~~~~~~~~~~~~~~~~~~~~~~~
\end{equation}

\begin{eqnarray}\label{17}
K_{-1,0}=\sum_{k=3}^{12}b_k^{1,0}(n,l,m)P_{-3,0}\left(\alpha_k^{1,0},\beta_k^{1,0},\gamma_k^{1,0}\right)+
(m+n+1)P_{-3,0}(3,0,1)+
~\nonumber~~~~~~~~~~~~\\
(m+3)P_{-3,0}(1,2,1)+\left[4l^2-2m+2l(2m+1)-2n(m+3)-3\right]P_{-3,0}(0,2,1)+
~\nonumber~~~~\\
\left[4l(m+3)-2m(n-1)-n(n+1)+3\right]P_{-3,0}(2,0,1),~~~~~~~~~~~~~~~~
\end{eqnarray}

\begin{eqnarray}\label{18}
K_{1,1}=\sum_{k=1}^{14}b_k^{1,0}(n,l,m)P_{-1,1}\left(\alpha_k^{1,0},\beta_k^{1,0},\gamma_k^{1,0}\right)+
2\left[P_{-1,0}\left(3,0,1\right)-P_{-1,0}\left(1,2,1\right)\right]+
~\nonumber~~~~~~\\
4(2l+m+n+3)\left[P_{-1,0}\left(0,2,1\right)-P_{-1,0}\left(2,0,1\right)\right],~~~~~~~~~~~~~~~~~~~~~~~~~~~
\end{eqnarray}

\begin{eqnarray}\label{19}
K_{1,2}=\sum_{k=1}^{14}b_k^{1,0}(n,l,m)P_{-1,2}\left(\alpha_k^{1,0},\beta_k^{1,0},\gamma_k^{1,0}\right)+
8\left[P_{-1,0}\left(0,2,1\right)-P_{-1,0}\left(2,0,1\right)\right]+
~\nonumber~~~~~~\\
4\left[P_{-1,1}\left(3,0,1\right)-P_{-1,1}\left(1,2,1\right)\right]+
8(2l+m+n+3)\left[P_{-1,1}\left(0,2,1\right)-P_{-1,1}\left(2,0,1\right)\right],~
\end{eqnarray}

\begin{eqnarray}\label{20}
K_{0,2}=\sum_{k=3}^{12}b_k^{1,0}(n,l,m)P_{-2,2}\left(\alpha_k^{1,0},\beta_k^{1,0},\gamma_k^{1,0}\right)+
(m+n+2)P_{-2,2}\left(3,0,1\right)+(m+2)P_{-2,2}\left(1,2,1\right)+
\nonumber~\\
2\left[l(2m+2l+3)-n(m+2)\right]P_{-2,2}\left(0,2,1\right)+\left[4l(m+2)-n(2m+n+3)\right]P_{-2,2}\left(2,0,1\right)+
\nonumber~~~~~~\\
8(2l+m+n+2)\left[P_{-2,1}\left(0,2,1\right)-P_{-2,1}\left(2,0,1\right)\right]+
4\left[P_{-2,1}\left(3,0,1\right)-P_{-2,1}\left(1,2,1\right)\right]+
~\nonumber~~~~~~~~\\
8\left[P_{-2,0}\left(0,2,1\right)-P_{-2,0}\left(2,0,1\right)\right]
.~~~~~~~~~~~~~~~~~~~~~~~~~~~~~~~~
\end{eqnarray}
The coefficients $b^{0,0},\alpha^{0,0},\beta^{0,0},\gamma^{0,0}$ and $b^{1,0},\alpha^{1,0},\beta^{1,0},\gamma^{1,0}$
can be found in Table \ref{T1}.

\section{Analytic expressions for the basic integrals.}\label{S2}

Our objective is to derive analytical expressions for the basic integral (\ref{11}).
To this end, first of all, changing the order of integration one obtains:
\begin{eqnarray}\label{31}
\mathcal{P}_{\imath,\jmath}(\nu,\ell,\mu)=\int_0^\infty e^{-s}s^{\nu}ds
\int_0^s t^{\ell}\left(s^2+t^2\right)^{\imath/2}\left( \ln \frac{s^2+t^2}{2}\right)^{\jmath} dt\int_t^s u^{\mu}du=
\nonumber~~~~~~\\
\frac{1}{\mu+1}\int_0^\infty e^{-s}s^\nu \left[s^{\mu+1}\mathcal{I}_{\imath,\jmath}^{(\ell)}(s)-
\mathcal{I}_{\imath,\jmath}^{(\ell+\mu+1)}(s)\right]ds,~~~~~~~~~~~~~~~~~~~~~~~~~~~~
\end{eqnarray}
where $\mu\geq0$ is integer, and
\begin{equation}\label{32}
\mathcal{I}_{\imath,\jmath}^{(p)}(s)\equiv\int_0^s t^p\left(s^2+t^2\right)^{\imath/2}\left( \ln \frac{s^2+t^2}{2}\right)^{\jmath} dt,
\end{equation}
with integer $p\geq 0$.

First, let us consider the simplest case of $\jmath=0$ (without logarithm).
For this case, integral (\ref{32}) can be presented in the form
\begin{equation}\label{33}
\mathcal{I}_{\imath,0}^{(p)}(s)=s^{p+\imath+1}\mathcal{F}_\imath(p),
\end{equation}
where the factor
\begin{equation}\label{34}
\mathcal{F}_{\imath}(p)=\frac{1}{p+1}~_2F_1\left(-\frac{\imath}{2},\frac{p+1}{2};\frac{p+3}{2};-1\right)
\end{equation}
includes the Gauss hypergeometric function $~_2F_1(a,b;c;z)$.
Notice, that formulas (\ref{33})-(\ref{34}) are valid for both positive and negative values of $\imath$.
For integer $p$ and $\imath \geq 0$ the hypergeometric function, presented in Eq.(\ref{34}),
can be reduced, if necessary, to combination of the elementary functions.
For the negative even $\imath$ this hypergeometric function can be cast, for example, to some combination of the digamma functions.

Inserting representation (\ref{33}) into Eq.(\ref{31}), and performing a trivial integration, one obtains:
\begin{eqnarray}\label{35}
\mathcal{P}_{\imath,0}(\nu,\ell,\mu)=\frac{1}{\mu+1}\int_0^\infty e^{-s}s^\nu \left[s^{\mu+1}\mathcal{I}_{\imath,0}^{(\ell)}(s)-\mathcal{I}_{\imath,0}^{(\ell+\mu+1)}(s)\right]ds=
\nonumber~~~~~~\\
\frac{(\imath+\nu+\ell+\mu+2)!}{\mu+1}
\left[\mathcal{F}_\imath(\ell)-\mathcal{F}_\imath(\ell+\mu+1) \right].
\end{eqnarray}
To treat the integral (\ref{31}) with integer $\jmath >0$, let us start with changing variables, $x=\left(s^2+t^2\right)/2$ in the integral (\ref{32}), what yields:
\begin{equation}\label{36}
\mathcal{I}_{\imath,\jmath}^{(p)}(s)=
\int_{s^2/2}^{s^2}\left(2x-s^2\right)^{(p-1)/2}\left(2x\right)^{\imath/2}\ln^\jmath x~dx.
\end{equation}

\subsection{Odd powers $p$}\label{S2a}

First, let us consider the case of odd $p=2q+1$ $(q=0,1,2,...)$. One obtains:
\begin{equation}\label{37}
\mathcal{I}_{\imath,\jmath}^{(2q+1)}(s)=
\int_{s^2/2}^{s^2}\left(2x-s^2\right)^{q}\left(2x\right)^{\imath/2}\ln^\jmath x~dx=
\sum_{k=0}^q
\left(\begin{array}{c} q \\ k \\
\end{array}\right)
2^{k+\imath/2}\left(-s^2\right)^{q-k}\mathcal{J}_{\jmath}^{(k+\imath/2)}(s),
\end{equation}
with
\begin{equation}\label{38}
\mathcal{J}_{\jmath}^{(\kappa)}(s)=\int_{s^2/2}^{s^2}x^{\kappa}\ln^\jmath x~dx.
\end{equation}
It is clear that $\left(\begin{array}{c} q \\ k \\\end{array}\right)$  are the binomial coefficients.

For integral (\ref{38}), \emph{Mathematica} \cite{MAT} gives:
\begin{equation}\label{39}
\mathcal{J}_{\jmath}^{(\kappa)}(s)=\frac{(-1)^\jmath}{(\kappa+1)^{\jmath+1}}
\left\{\Gamma \left[\jmath+1,-2(\kappa+1)\ln s\right]-\Gamma\left[\jmath+1,(\kappa+1)\ln \left(2/s^2\right)\right]
\right\},
\end{equation}
where $\Gamma(a,z)$ is the incomplete gamma function. It was examined numerically, that representation (\ref{39}) is valid
(at least) for integer $\jmath\geq0$, real $s>0$, and real $\kappa$, excluding only $\kappa=-1$.
Using the well-known relation
\begin{equation}\label{40}
\Gamma \left(\jmath+1,z\right)=\jmath!\exp(-z)\sum_{n=0}^\jmath \frac{z^n}{n!},
\end{equation}
for integer $\jmath\geq0$, one obtains instead of (\ref{39}):
\begin{eqnarray}\label{41}
\mathcal{J}_{\jmath}^{(\kappa)}(s)=\frac{\jmath!(-1)^{\jmath+1}}{2^{\kappa+1}(\kappa+1)^{\jmath+1}}s^{2(\kappa+1)}
\sum_{n=0}^\jmath \frac{(\kappa+1)^n}{n!}\left[\left(\ln\frac{2}{s^2}\right)^n+(-2)^{n+1}2^\kappa \left(\ln s\right)^n\right]=
\nonumber~~~~~~\\
\frac{\jmath!(-1)^{\jmath+1}}{2^{\kappa+1}(\kappa+1)^{\jmath+1}}s^{2(\kappa+1)}
\sum_{n=0}^\jmath c_n^{(\jmath)}(\kappa)\ln^n s,~~~~~~~~~~~~~~~~~~~~~~~
\end{eqnarray}
where after some transforms and simplifications one obtains the following two equivalent representations for the coefficients:
\begin{eqnarray}\label{42}
c_n^{(\jmath)}(\kappa)=\frac{2^{n+\kappa+1}(-\kappa-1)^n}{n!}
\left[ \frac{\Gamma\left(\jmath-n+1,(\kappa+1)\ln 2\right)}{(\jmath-n)!}-1\right]=
\nonumber~~~~~~~~~~~~~~~~~\\
\frac{2^{n}(-\kappa-1)^n}{n!}
\left[\sum_{m=0}^{\jmath-n}\frac{(\kappa+1)^m \ln^m 2}{m!}-2^{\kappa+1}
\right].
\end{eqnarray}
The same result could be certainly obtained by using relation (1.6.1.6) \cite{PRU}.

Substituting the latter result into Eq.(\ref{37}), one obtains:
\begin{equation}\label{43}
\mathcal{I}_{\imath,\jmath}^{(2q+1)}(s)=
\sum_{k=0}^q
\left(\begin{array}{c} q \\ k \\\end{array}\right)
2^{k+\imath/2}\left(-s^2\right)^{q-k}\mathcal{J}_{\jmath}^{(k+\imath/2)}(s)=
(-1)^{\jmath+q+1}2^{\imath/2-1}s^{2q+\imath+2}\sum_{n=0}^\jmath \mathcal{B}_{n}^{(\imath,\jmath)}(q)\ln^n s,
\end{equation}
where
\begin{equation}\label{44}
\mathcal{B}_{n}^{(\imath,\jmath)}(q)=\left(\begin{array}{c} \jmath \\ n \\ \end{array}\right)
\sum_{k=1}^{q+1}
\left( \begin{array}{c} q \\ k-1 \\\end{array}\right) \frac{(-2)^{k+n}}{\left(k+\imath/2\right)^{\jmath-n+1}}
\left[(\jmath-n)!-\Gamma\left(\jmath-n+1,(k+\imath/2)\ln 2\right)\right].
\end{equation}
For the case of $\kappa=-1$, (see, (1.6.1.7) \cite{PRU}) integral (\ref{38}) reads
\begin{equation}\label{45}
\mathcal{J}_\jmath^{(-1)}(s)=\frac{2^{\jmath+1}}{\jmath+1}
\left[\ln^{j+1}s-\left(\ln s-\frac{\ln 2}{2}\right)^{\jmath+1}\right]=
-\frac{(-\ln 2)^{\jmath+1}}{\jmath+1}
\sum_{n=0}^{\jmath}\left(\begin{array}{c} \jmath+1 \\ n \\\end{array}\right)
\left(-\frac{2}{\ln 2}\right)^n \ln^n s.
\end{equation}
Integral (\ref{45}) is included into the right-hand side of Eq.(\ref{37}) only for $\imath=-2$, whence:
\begin{eqnarray}\label{46}
\mathcal{I}_{-2,\jmath}^{(2q+1)}(s)=\frac{1}{2}\left(-s^2\right)^q \mathcal{F}_\jmath^{(-1)}+
\sum_{k=1}^q \left(\begin{array}{c} q \\ k \\
\end{array}\right)
2^{k-1}\left(-s^2\right)^{q-k}\mathcal{J}_{\jmath}^{(k-1)}(s)=
\nonumber~~~~~~\\
(-1)^{\jmath+q+1}s^{2q}\sum_{n=0}^\jmath \mathcal{B}_{n}^{(-2,\jmath)}(q)\ln^n s.~~~~~~~~~~~~~
\end{eqnarray}
Thus, for this case we obtain the same representation (\ref{43}) (for $\imath=-2$), but with coefficient:
\begin{equation}\label{47}
\mathcal{B}_{n}^{(-2,\jmath)}(q)=(-2)^{n-1}\left(\begin{array}{c} \jmath \\ n \\\end{array}\right)\left\{
\frac{(\ln 2)^{\jmath-n+1}}{\jmath-n+1}+
\sum_{k=1}^q
\left(\begin{array}{c} q \\ k \\\end{array}\right)\frac{(-2)^{k}}{k^{\jmath-n+1}}
\left[ (\jmath-n)!-\Gamma \left(\jmath-n+1,k \ln 2 \right)\right]\right\}.
\end{equation}
The incomplete gamma functions included into Eqs.(\ref{44}), (\ref{47}), if necessary, can be expressed in form (\ref{40}).

\subsection{Even powers $p$}\label{S2b}

It is clear that for even $p=2q$ integral (\ref{36}) can be presented in the form:
\begin{equation}\label{50}
\mathcal{I}_{\imath,\jmath}^{(2q)}(s)=
I_{\imath,\jmath}^{(q)}\left(s,s^2\right)-
I_{\imath,\jmath}^{(q)}\left(s,\frac{s^2}{2}\right),
\end{equation}
where
\begin{equation}\label{51}
I_{\imath,\jmath}^{(q)}(s,x)=
\int\left(2x-s^2\right)^{q-1/2}(2x)^{\imath/2}\ln^\jmath x~dx.
\end{equation}
is the indefinite integral pertaining to the case.

Using a variety tools of \emph{Mathematica}-9, we have derived the following representation for the more general
indefinite integral:
\begin{eqnarray}\label{52}
\int (2x)^{\mu-1}(2x-y)^\lambda (\ln x)^n dx=\frac{(2x)^\mu}{2\mu}\left(-\frac{1}{y}\right)^{-\lambda}
\left[(\ln x)^n~_2F_1\left(\mu,-\lambda;\mu+1;\frac{2x}{y}\right)+
\right.
\nonumber~~~~~~~~~~\\
\left.
\sum_{m=1}^n \left(-\frac{1}{\mu}\right)^m \frac{n!}{(n-m)!}(\ln x)^{n-m}
~_{m+2}F_{m+1}\left(\mu,...,\mu,-\lambda;\mu+1,...,\mu+1;\frac{2x}{y}\right)
\right],~~~~~~~~~~
\end{eqnarray}
where $~_nF_m\left(a_1,...,a_n;b_1,...,b_m;z\right)$ is generalized hypergeometric function. It is seen, that
formula (\ref{52}) cannot be applied for $\mu=0$. In this specific case, one obtains:
\begin{eqnarray}\label{53}
\int (2x)^{-1}(2x-y)^\lambda (\ln x)^n dx=\frac{(2x)^\lambda}{2\lambda}
\left[(\ln x)^n~_2F_1\left(-\lambda,-\lambda;1-\lambda;\frac{y}{2x}\right)+
\right.
\nonumber~~~~~~~~~~\\
\left.
\sum_{m=1}^n \left(-\frac{1}{\lambda}\right)^m \frac{n!}{(n-m)!}(\ln x)^{n-m}
~_{m+2}F_{m+1}\left(-\lambda,-\lambda,...,-\lambda;1-\lambda,...,1-\lambda;\frac{y}{2x}\right)
\right].~~~~~~~~~~
\end{eqnarray}
Notice that we could not find representations (\ref{52}), (\ref{53}) in scientific literature.

It is clear that Eq.(\ref{53}) cannot be applied for integer $\lambda\geq0$. In this case, one should make use the binomial expansion and then to apply formula (1.6.1.6-7) from handbook \cite{PRU}. For the considered integral (\ref{36}) this case was treated in the previous section.

Using Eq.(\ref{52}), one can present integral (\ref{50}) in the form:
\begin{equation}\label{54}
\mathcal{I}_{\imath,\jmath}^{(2q)}(s)=\textbf{i}(-1)^q s^{2q+\imath+1}\sum_{n=0}^\jmath
\mathcal{A}_n^{(\imath,\jmath)}(q)(\ln s)^n,
\end{equation}
where the coefficients
\begin{eqnarray}\label{55}
\mathcal{A}_n^{(\imath,\jmath)}(q)=\frac{2^n \jmath !(-1)^{\jmath-n}}{(\imath+2)n!}
\left[-\frac{\Gamma\left(\frac{\imath}{2}+2\right)\Gamma\left(q+\frac{1}{2}\right)(\ln 2)^{\jmath-n}}
{\Gamma\left(\frac{\imath+3}{2}+q\right)(\jmath-n)!}+
\right.
\nonumber~~~~~~~~~~\\
2^{\imath/2+1}\left(\frac{2}{\imath+2}\right)^{\jmath-n}~_{\jmath-n+2}F_{\jmath-n+1}
\left(\frac{\imath}{2}+1,...,\frac{\imath}{2}+1,\frac{1}{2}-q;\frac{\imath}{2}+2,...,\frac{\imath}{2}+2;2\right)-
\nonumber~~~~~~~~~~\\
\left.
\sum_{m=n}^{\jmath-1}\frac{\left(\frac{2}{\imath+2}\right)^{\jmath-m}(\ln 2)^{m-n}}{(m-n)!}
~_{\jmath-m+2}F_{\jmath-m+1}
\left(\frac{\imath}{2}+1,...,\frac{\imath}{2}+1,\frac{1}{2}-q;\frac{\imath}{2}+2,...,\frac{\imath}{2}+2;1\right)
\right]~~~~~~
\end{eqnarray}
are correct for $0\leq n\leq \jmath-1$. For $n=\jmath$ we get:
\begin{equation}\label{56}
\mathcal{A}_\jmath^{(\imath,\jmath)}(q)=-2^{\jmath-1}\mathcal{Q}(\imath,q)
\end{equation}
where, using, e.g., (7.3.1.28) \cite{PRU3}, one can write down:
\begin{equation}\label{57}
\mathcal{Q}(\imath,q)=
\frac{\Gamma\left(\frac{\imath}{2}+1\right)\Gamma\left(q+\frac{1}{2}\right)}{\Gamma\left(\frac{\imath+3}{2}+q\right)}
-B_2\left(\frac{\imath}{2}+1,q+\frac{1}{2}\right).
\end{equation}
Here $\Gamma(z)$ is the Euler gamma function, and $B_z(b,a)$ is the incomplete beta function.
The last expression can be simplified and presented as a linear combination of rational numbers, $\sqrt{2}$ and $\emph{arccos}(\sqrt{2})$. In particular, one obtains
\begin{equation}\label{58}
\mathcal{Q}(\imath,q)=\frac{\textbf{i}(-1)^q(\imath/2)!2^{\imath/2+1}}{(2q+\imath+1)!!}
\sum_{k=0}^{\imath/2}\frac{(2q+2k-1)!!}{k!},
\end{equation}
for even $\imath$, and
\begin{eqnarray}\label{59}
\mathcal{Q}(\imath,q)=\frac{(2q-1)!!}{(q+(\imath+1)/2)!}\Bigg\{\frac{\imath !!}{2^{q+\imath/2}}\Bigg[
\sqrt{2}\arccos(\sqrt{2})+\textbf{i}\sum_{k=0}^{(\imath-1)/2}\frac{2^{\imath-2k}((\imath-1)/2-k)!}{(\imath-2k)!!}
\Bigg]+
\nonumber~~~~\\
2\textbf{i}\sum_{k=0}^{q-1}\frac{(-1)^{q-k}2^{\imath/2-k}(q+(\imath-1)/2-k)!}{(2q-2k-1)!!}
\Bigg\},~~~~~~~~~~~~~~~~~~~~~~~~
\end{eqnarray}
for odd values of $\imath$.

It is seen, that expression (\ref{55}) is not correct for $\imath=-2$. This specific case corresponds to $\mu=0$ in Eq.(\ref{52}). Therefore, in this case we used Eq.(\ref{53}) in order to present integral (\ref{50}) in the same form (\ref{54}), but with coefficients
\begin{eqnarray}\label{60}
\mathcal{A}_n^{(-2,\jmath)}(q)=\frac{\textbf{i}2^{n-1}\jmath!}{n!}(-1)^{q+\jmath-n+1}\Bigg\{
\frac{\pi(-1)^q(\ln 2)^{\jmath-n}}{(\jmath-n)!}+
\nonumber~~~~~~~~~~\\
2^{q-\frac{1}{2}}\left(\frac{2}{2q-1}\right)^{\jmath-n+1}
~_{\jmath-n+2}F_{\jmath-n+1}
\left(\frac{1}{2}-q,...,\frac{1}{2}-q;\frac{3}{2}-q,...,\frac{3}{2}-q;\frac{1}{2}\right)-
\nonumber~~~~~~~~~~\\
\sum_{m=n}^{\jmath-1}\frac{\left(\frac{2}{2q-1}\right)^{\jmath-m+1}(\ln 2)^{m-n}}{(m-n)!}
~_{\jmath-m+2}F_{\jmath-m+1}
\left(\frac{1}{2}-q,...,\frac{1}{2}-q;\frac{3}{2}-q,...,\frac{3}{2}-q;1\right)
\Bigg\},
\end{eqnarray}
that are correct for $0\leq n\leq\jmath-1$. For $n=\jmath$, one can use representation (\ref{56}),
but with function
\begin{equation}\label{61}
\mathcal{Q}(-2,q)=\textbf{i}\left[\pi-
(-1)^q B_{\frac{1}{2}}\left(\frac{1}{2}-q,\frac{1}{2}+q\right)
\right]=
\textbf{i}\left\{\pi-
\frac{(-1)^q}{2} \left[\boldsymbol{\psi}\left(\frac{3-2q}{4}\right)-\boldsymbol{\psi}\left(\frac{1-2q}{4}\right)\right]
\right\},
\end{equation}
which, in its turn, can be expressed in terms of rational numbers by using the relation:
\begin{equation}\label{62}
\boldsymbol{\psi}\left(\frac{3-2q}{4}\right)-\boldsymbol{\psi}\left(\frac{1-2q}{4}\right)=
\left\{ \begin{array}{c}\mathlarger{\pi+8\sum_{k=0}^{(q-2)/2}\frac{1}{(4k+1)(4k+3)}},~~~~~~~~~~~~~for~even~~q \\
 \mathlarger{-\pi-\frac{4}{2q+1}-8\mathlarger{\sum}_{k=0}^{(q-1)/2}\frac{1}{(4k+1)(4k+3)}},~~for~ odd~~ q. \\\end{array}\right.
\end{equation}
Notice that it is possible to apply formula (7.10.2.6) \cite{PRU3}, in order to express
the generalized hypergeometric function with an argument of unity (see, Eqs.(\ref{55}), (\ref{60})) through
the Euler gamma function, $\Gamma(z)$ and its derivatives.
However, we believe that the following representation
\begin{eqnarray}\label{63}
~_{n+2}F_{n+1}\left(a,...,a,b;a+1,...,a+1;1\right)=
\nonumber~~~~~~~~~~~~~~~~~~~~~~~~~\\
\frac{(-1)^n a^{n+1}}{n!}B(a,1-b)
Y_n\left(\boldsymbol{\psi}(a)-\boldsymbol{\psi}(1+a-b),...,\boldsymbol{\psi}^{(n-1)}(a)-\boldsymbol{\psi}^{(n-1)}(1+a-b)\right),~~~
\end{eqnarray}
given in \cite{KK} (see, also references therein), is the most simple and effective. Here, $\boldsymbol{\psi}^{(n)}(z)=d^n\boldsymbol{\psi}(z)/dz^n$ is the $n^{th}$ derivative of the digamma function $\boldsymbol{\psi}(z)$, and $B(x,y)=\Gamma(x)\Gamma(y)/\Gamma(x+y)$ is the Euler  beta function.
The $n^{th}$ complete Bell polynomial $Y_n(x_1,...,x_n)=\sum_{k=1}^nY_{n,k}(x_1,x_2,...,x_{n-k+1})$ presents a sum of the
partial Bell polynomials $Y_{n,k}$. The first few complete Bell polynomials are:
\begin{eqnarray}\label{64}
Y_1(x_1)=x_1,
\nonumber~~~~~~~~~~~~~~~~~~~~~~~~~~~~~~~~~~~~~~~~~~~~~~~~~~~~~~~~~~~~~~~~~\\
Y_2(x_1,x_2)=x_1^2+x_2,
\nonumber~~~~~~~~~~~~~~~~~~~~~~~~~~~~~~~~~~~~~~~~~~~~~~~~~~~~~~\\
Y_3(x_1,x_2,x_3)=x_1^3+3x_1x_2+x_3,
\nonumber~~~~~~~~~~~~~~~~~~~~~~~~~~~~~~~~~~~~~~~\\
Y_4(x_1,x_2,x_3,x_4)=x_1^4+6x_1^2x_2+3x_2^2+4x_1x_3+x_4.~~~~~~~~~~~~~~~~
\end{eqnarray}
It is seen from (\ref{64}) that the right-hand side of Eq.(\ref{63}) presents the linear combination of the terms of the form:
\begin{equation}\label{65}
\left[\boldsymbol{\psi}^{(m_1)}(a)-\boldsymbol{\psi}^{(m_1)}(1+a-b)\right]^{\ell_1}\left[\boldsymbol{\psi}^{(m_2)}(a)-\boldsymbol{\psi}^{(m_2)}(1+a-b)\right]^{\ell_2},
\end{equation}
where $(m_1+1)\ell_1+(m_2+1)\ell_2$ equals the order $n$ of the considered Bell polynomial $Y_n$.
According to Eq.(\ref{60}), we are interested in the case of $a=b=1/2-q$.
For this (simplest) case one obtains:
\begin{equation}\label{66}
\boldsymbol{\psi}^{(m)}\left(\frac{1}{2}-q\right)=
\left\{ \begin{array}{c}(2^{m+1}-1)\boldsymbol{\psi}^{(m)}(1)+m!2^{m+1}\mathlarger{\sum_{k=1}^q\frac{1}{(2k-1)^{m+1}}},
~~~~~~~~m>0 \\
\mathlarger{\sum_{k=1}^{q-1}\frac{1}{k}+\sum_{k=q}^{2q-1}\frac{2}{k}}-\ln 4+\boldsymbol{\psi}(1),
~~~~~~~~~~~~~~~~~~~~~~~~~~~~~~~m=0 \\\end{array}\right.
\end{equation}
\begin{equation}\label{67}
\boldsymbol{\psi}^{(m)}(1)=(-1)^{m+1}m!\boldsymbol{\zeta}(m+1),~~~~~~~~~~~~~~~~~~~~~~~~~~~~~~~~~~~(m>0)
\end{equation}
with $\boldsymbol{\psi}(1)=-\boldsymbol{\gamma}$, where $\boldsymbol{\gamma}$ is Euler's constant, and $\boldsymbol{\zeta}(z)$ is the Riemann zeta function.

In accordance with Eq.(\ref{55}), one should consider the case of $a=\imath/2+1,~b=1/2-q$.
For this case one obtains:
\begin{eqnarray}\label{68}
\boldsymbol{\psi}^{(m)}\left(\frac{\imath}{2}+1\right)-\boldsymbol{\psi}^{(m)}\left(\frac{\imath+3}{2}+q\right)=
\nonumber~~~~~~~~~~~~~~~~~~~~~~~~~~~~~~~~~~~~~~~~~\\
k_0\left\{\boldsymbol{\psi}^{(m)}\left(\frac{1}{2}\right)-\boldsymbol{\psi}^{(m)}\left(1\right)+(-1)^m m!\left[2^{m+1}\sum_{k=0}^{k_1}\frac{1}{(2k+1)^{m+1}}-\sum_{k=0}^{k_2}\frac{1}{(k+1)^{m+1}}\right]\right\},~~~~~~~~~~
\end{eqnarray}
where, one should put $k_0=1,~k_1=(\imath-1)/2,~k_2=k_1+q$, for odd values of $\imath$, and
$k_0=-1,~k_1=\imath/2+q,~k_2=\imath/2-1$, for even $\imath$. Representation for $\boldsymbol{\psi}^{(m)}(1/2)$ can be found,
putting $q=0$ in Eq.(\ref{66}).

\subsection{Final derivation}\label{S2c}

Thus, Eqs.(\ref{33}), (\ref{43}) and (\ref{54}) enable us to present integrals (\ref{32}) in the form
\begin{equation}\label{70}
\mathcal{I}_{\imath,\jmath}^{(p)}(s)=s^{p+\imath+1}\sum_{n=0}^\jmath \mathcal{C}_n^{(\imath,\jmath)}(p)(\ln s)^n,
\end{equation}
where $\mathcal{C}$-coefficients can be expressed through $\mathcal{A}$-coefficients for even $p$, and through
$\mathcal{B}$-coefficients for odd $p$. Inserting representation (\ref{70}) into Eq.(\ref{31}), one obtains:
\begin{equation}\label{71}
\mathcal{P}_{\imath,\jmath}(\nu,\ell,\mu)=\frac{1}{\mu+1}\sum_{n=0}^\jmath
\left[\mathcal{C}_n^{(\imath,\jmath)}(\ell)-\mathcal{C}_n^{(\imath,\jmath)}(\ell+\mu+1)\right]
X_n(\nu+\mu+\ell+\imath+2),~~~~(\mu\geq0)
\end{equation}
where (see, (2.6.21.1) \cite{PRU})
\begin{equation}\label{72}
X_n(\alpha)=\int_0^\infty e^{-s}s^\alpha (\ln s)^n ds=\frac{d^n}{d\alpha^n}\Gamma(\alpha+1).~~~~~~~~~~~~(\alpha>-1)
\end{equation}
The most convenient and effective to express the integral (\ref{72}) through the complete Bell polynomials, what gives:
\begin{equation}\label{73}
 X_n(\alpha)=\Gamma(\alpha+1)Y_n\left(\boldsymbol{\psi}(\alpha+1),...,\boldsymbol{\psi}^{(n-1)}(\alpha+1)\right).~~~~~~~~~~~(n\geq1)
\end{equation}
In the trivial case of $n=0$, one should simply set $Y_0=1$ .

Thus, for integer $\alpha$, one finally obtains instead of (\ref{71}):
\begin{eqnarray}\label{74}
\mathcal{P}_{\imath,\jmath}(\nu,\ell,\mu)=\frac{(\nu+\mu+\ell+\imath+2)!}{\mu+1}\times
\nonumber~~~~~~~~~~~~~~~~~~~~~~~~~~~~~~~~~~~~~~~~~\\
\sum_{n=0}^{\jmath}
\left[\mathcal{C}_n^{(\imath,\jmath)}(\ell)-\mathcal{C}_n^{(\imath,\jmath)}(\ell+\mu+1)\right]
Y_n\left(\boldsymbol{\psi}(\nu+\mu+\ell+\imath+3),...,\boldsymbol{\psi}^{(n-1)}(\nu+\mu+\ell+\imath+3) \right).~~~
\end{eqnarray}
Notice that for integer argument, the following expression holds:
\begin{equation}\label{75}
\boldsymbol{\psi}^{(n)}(m+1)=(-1)^n n!\left[-\boldsymbol{\zeta}(n+1)+\sum_{k=1}^m\frac{1}{k^{n+1}}\right].
\end{equation}

\section{Numerical results and conclusions}\label{S4}

In this section we present the results of testing the basis (\ref{6}) containing $\ln^j r$ with $j=0,1,2$.
Notice that in the simplest case one can use the individual (single) basis functions of the form (\ref{6}) to solve the generalized eigenvalue equation (\ref{8}). However, it is more effective to make use of the composite basis including basis functions build in the form of linear combination of the individual ones.
Therefore, we apply the composite basis function
\begin{equation}\label{76}
\phi_1=\exp\left(-\frac{s}{2}\right)\left[1-\left(\delta Z-\frac{1}{2}\right)s+\frac{\delta}{2}u-\frac{Z(\pi-2)}{3\pi}\delta^2(r^2-u^2)\ln r \right],
\end{equation}
that yields the exact leading terms of the Fock series expansion (\ref{1}) for arbitrary variational parameter $\delta$.
In particular, using of the basis function (\ref{76}) enables us to obtain the total two-electron wave function with the exact series expansion $\Psi\simeq 1+r \psi_{10}+r^2 \psi_{21}\ln r$ (see, e.g., \cite{GAM,FOR,MYE}).
For the most correct approximation of the term $r^2 \psi_{20}$ of the Fock expansion (\ref{1}), we used basis functions
$\phi_k=\exp(-s/2)\omega(s,t,u,r)$ with polynomial factor $\omega=sr,ur,su,s^2,u^3/r,t^2,u^2$ (see, e.g., \cite{FRP, ES1}) for $k=2,...,8$. To describe the term $r^3\psi_{30}$, the polynomials $\omega=sr^2,ur^2,s^2u,s^3,u^3,t^2 r,r(r^2-u^2)$ (see, e.g., \cite{ES1}) were used for $k=9,...,15$. And at last, we used the exact representation
$\omega \equiv r^3\psi_{31}\ln r=\left[6(Z s-u)(r^2-u^2)-u^3\right]\ln r$, for k=16 \cite{GAM}.
This set of 16 composite basis functions was extended with the help of 196, 201 and 195 individual basis functions
of the form (\ref{6}) with $n+2l+m+i\leq 6$, for the atom of helium, ions of hydrogen and lithium, respectively.
For selection of these functions we employed a scheme which is very closed to the one used by Frankowski \cite{FRP},
when the negative values of $n$ were admitted only for $j=0$, and the other integer indexes were non-negative.

The basis described above ($LR$-basis in what follows) produces the following ground state energies: -0.527 751 015 7 (-0.527 751 016 541) for $H^{-}$; -2.903 724 377 023 (-2.903 724 377 034) for $He$; -7.279 913 412 663 (-7.279 913 412 669 2) for $Li^{+}$. In parentheses, the corresponding results obtained with using the basis of Laguerre functions \cite{LEZ2, LEZ3} ($LF$-basis in what follows) are shown for comparison. Note that the size of the $LF$-basis is equal to 2856. It is seen that the $LR$-basis, which size is more that 10 times smaller, can produce the energies that are very close to the ones produced by the $LF$-basis.

\begin{figure}
\begin{center}
\subfloat[]{\epsfxsize=7.5cm\epsfbox{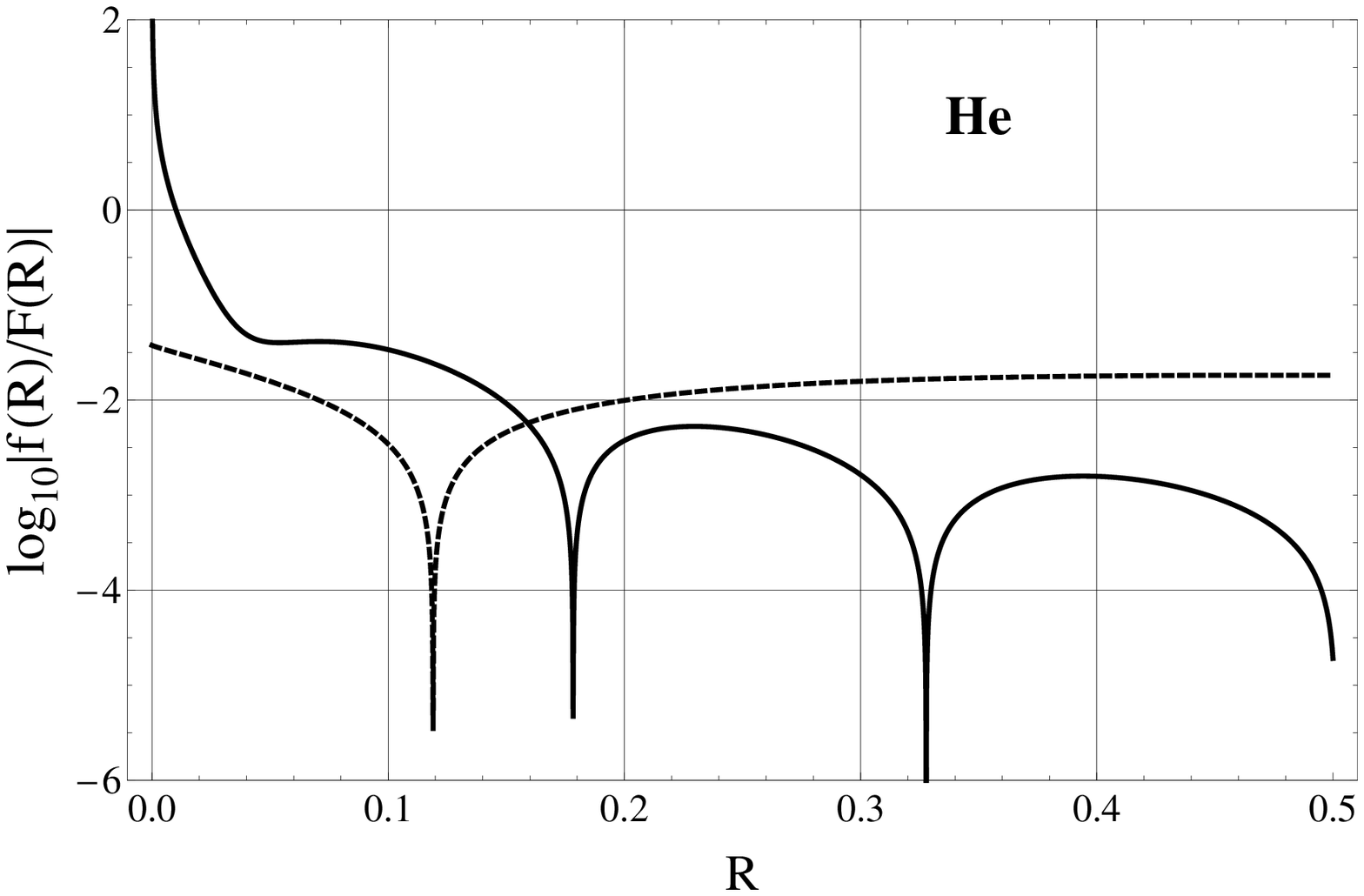}}
\subfloat[]{\epsfxsize=7.5cm\epsfbox{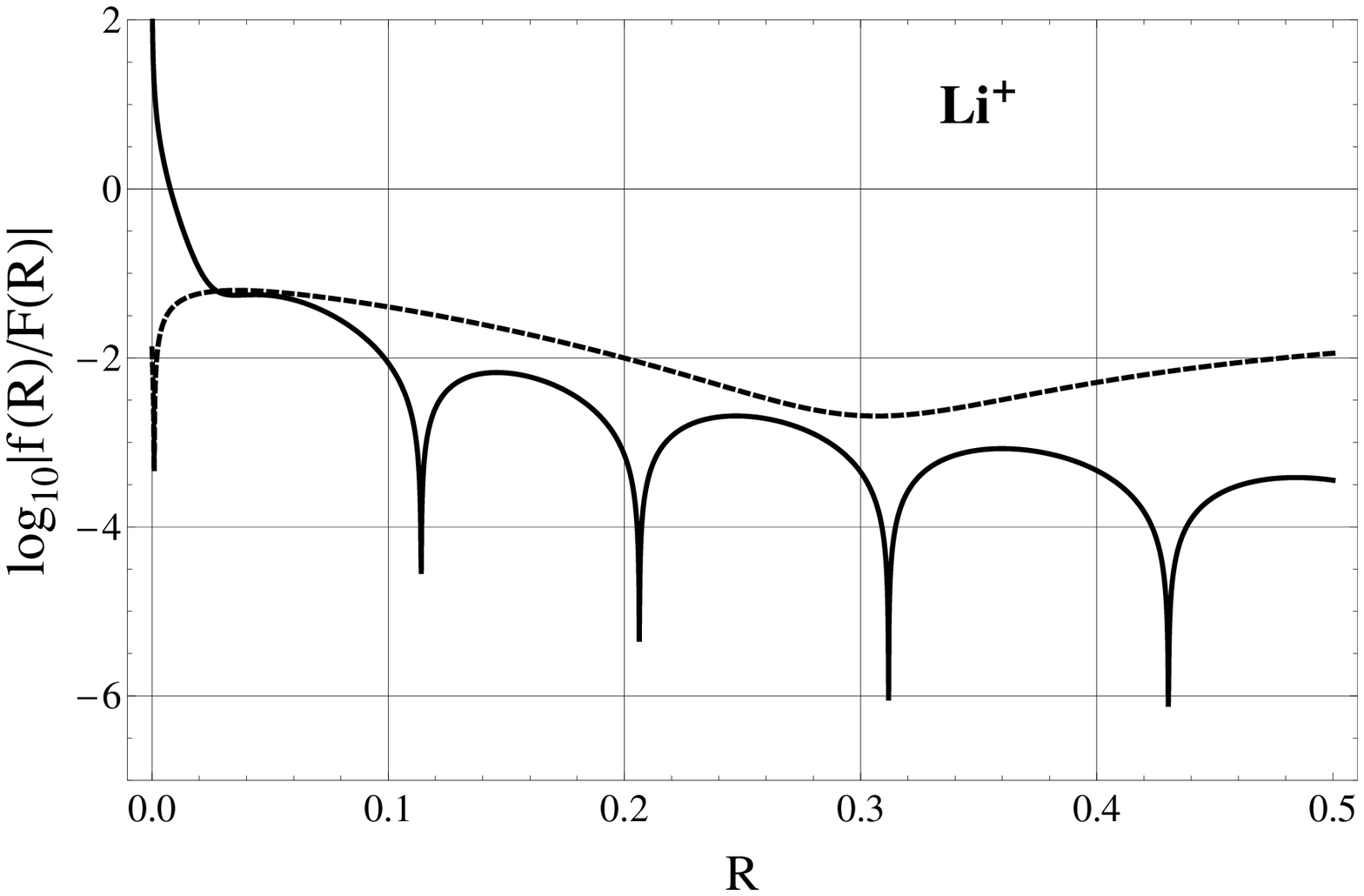}}
\caption{ Estimation of the accuracy $\log_{10}\left|f(R)/F(R) \right|$, of solution $F(R)=\Psi(R,0,R)$ to the Schr\"{o}dinger equation (\ref{7a}) at the electron-nucleus coalescence line. Solid and dash lines present the results obtained by using the $LF$-and $LR$-basis, respectively.}
\label{F1}
\end{center}
\end{figure}

\begin{figure}
\begin{center}
\subfloat[]{\epsfxsize=7.5cm\epsfbox{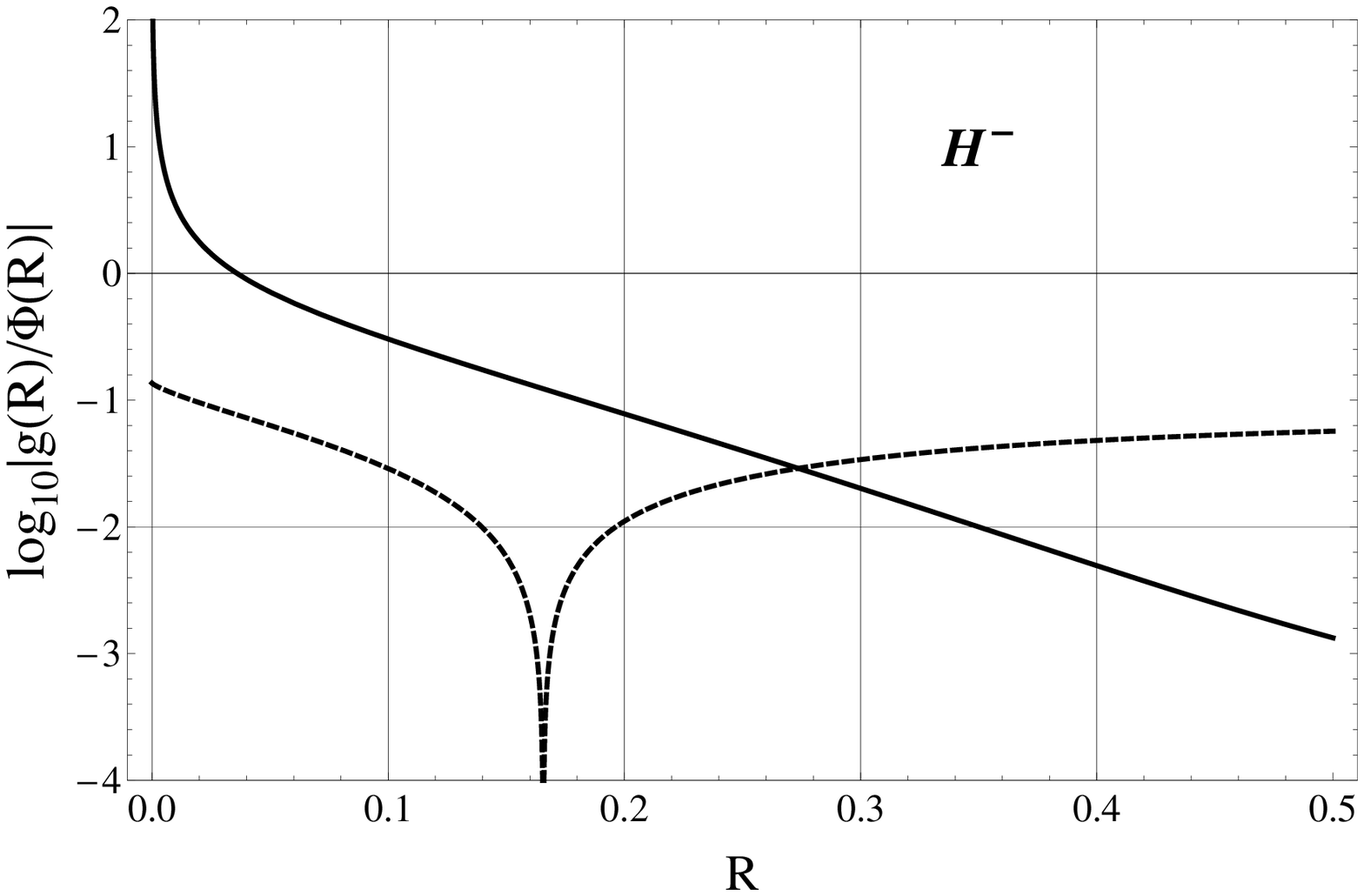}}
\subfloat[]{\epsfxsize=7.5cm\epsfbox{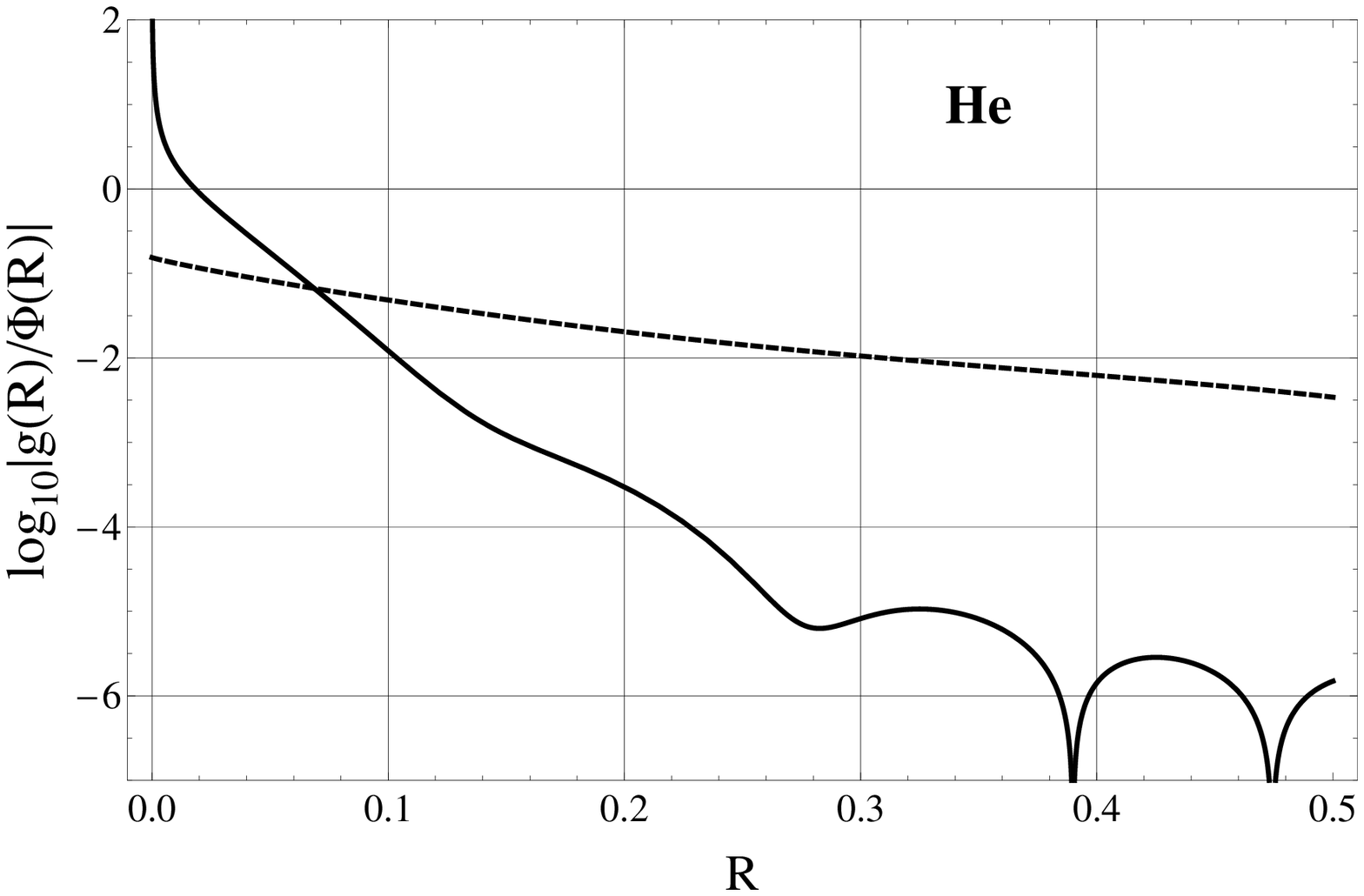}}\\
\subfloat[]{\epsfxsize=7.5cm\epsfbox{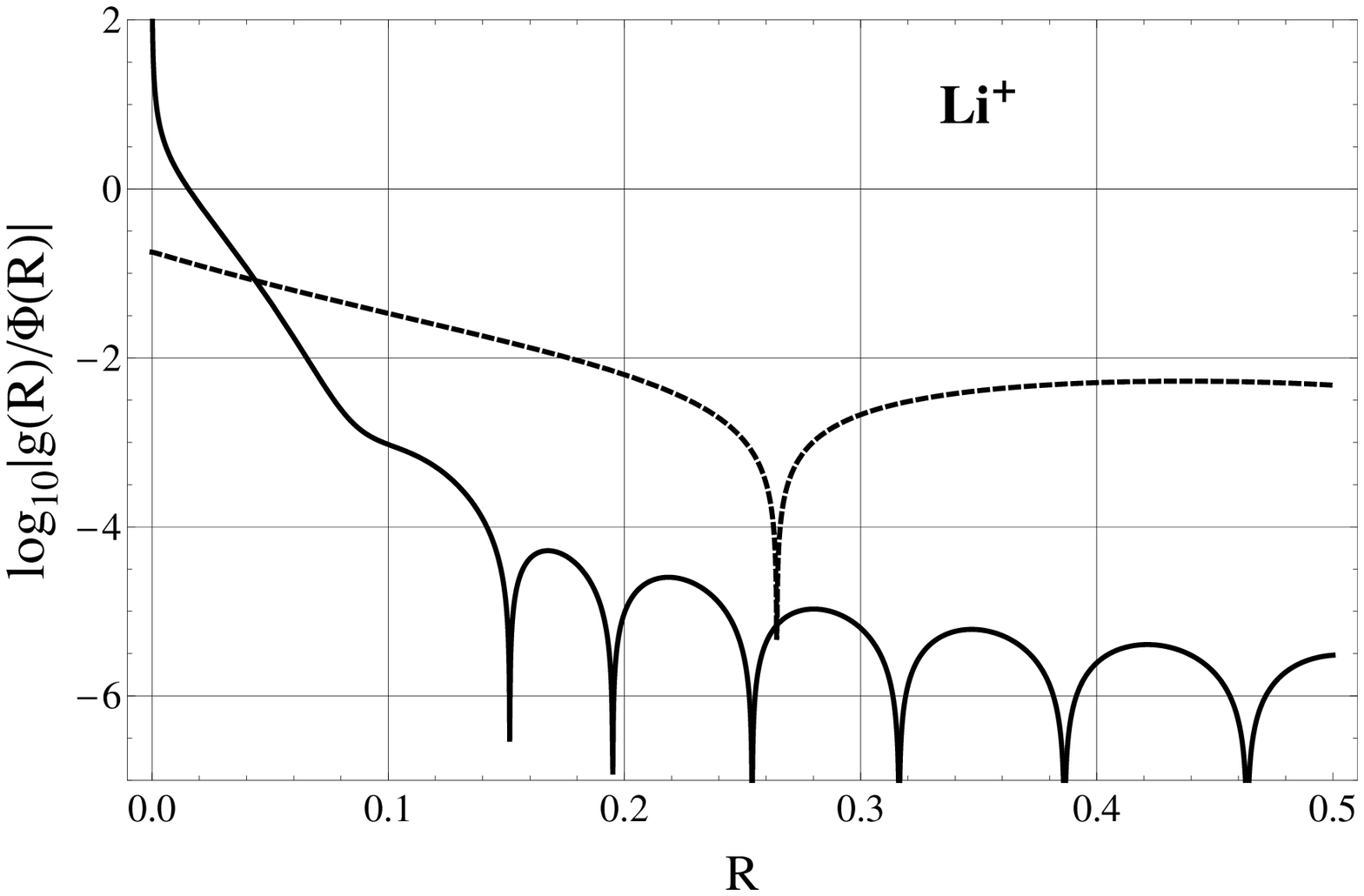}}
\caption{Estimation  of the accuracy $\log_{10}\left|g(R)/\Phi(R) \right|$, of solution $\Phi(R)=\Psi(R,R,0)$ to the Schr\"{o}dinger equation (\ref{7a}) at the electron-electron coalescence line.  Solid and dash lines present the results obtained by using the $LF$-and $LR$-basis, respectively.}
\label{F2}
\end{center}
\end{figure}

It was emphasized in Introduction that, first of all, we are interested in obtaining the two-electron wave function with correct behavior near the triple-coalescence point.
The simplest path to this point runs along the two-particle coalescence lines.
Remind that there are two kinds of the two-particle coalescences in the two-electron atomic systems.
Putting $r_1=r_{12}=R~(r_2=0)$ in the left-hand side of the Schr\"{o}dinger equation (\ref{7a}), one obtains the one-dimensional equation $f(R)=0$, whereas the wave function $\Psi(r_1,r_2,r_{12})$ transforms into $F(R)=\Psi(R,0,R)$.
This particular case corresponds to the formation of the electron-nucleus coalescence.
Putting $r_1=r_2=R~(r_{12}=0)$ in the left-hand side of the Schr\"{o}dinger equation (\ref{7a}), one obtains the equation $g(R)=0$, whereas the wave function $\Psi(r_1,r_2,r_{12})$ transforms into $\Phi(R)=\Psi(R,R,0)$, what corresponds to the formation of the electron-electron coalescence.
Estimation of the accuracy of solution to Eq.(\ref{7a}) can be done by means of the functions $\log_{10}\left|f(R)/F(R)\right|$ and $\log_{10}\left|g(R)/\Phi(R)\right|$,
which are displayed on Figs.\ref{F1} and \ref{F2}, respectively, for the first three atoms/ions of the helium-like sequence.
The results for the $LF$-basis and $LR$-basis are drawn by a solid line and a dash line, respectively.
We do not present the electron-nucleus coalescence for the negative ion of hydrogen, because in this case a questionable situation arises. We have in mind an extinction of the Coulomb interaction between the remote electron and the coalesced pair of the second electron and the hydrogen nucleus, because of vanishing of the total charge of the coalesced pair.
Both Fig.\ref{F1} and Fig.\ref{F2} show a significant increase in the accuracy of the ground state wave function,
calculated with using the $LR$-basis, near the origin $R=0$ (triple-coalescence point) at the two-particle coalescence lines.

It is important that calculations of the Hamiltonian matrix elements according to the analytic formulas presented in the previous sections take not more than 1 minute on a PC with a "i7-3770K
CPU@4.68 GHz" processor. We have in mind that all calculations were performed with the help of \emph{Mathematica} codes. Note that all of the special functions, presented in the previous sections, are the built-in \emph{Mathematica} programming units.

\section{Acknowledgment}

This work was supported by the Israel Science Foundation grant 954/09.

\newpage

\begin{table}
\begin{center}
\caption{Coefficients required to calculate the kinetic energy matrix according to Eqs.(\ref{14})-(\ref{20}).}
\begin{tabular}{|l||c|c|c|c|}
\hline
{\footnotesize $k$} & \multirow{1}{*}{{\footnotesize $b_{k}^{0,0}(n,l,m)$}} & {\footnotesize $\{\alpha_{k}^{0,0},\beta_{k}^{0,0},\gamma_{k}^{0,0}\}$} & {\footnotesize $b_{k}^{1,0}(n,l,m)$} & {\footnotesize $\{\alpha_{k}^{1,0},\beta_{k}^{1,0},\gamma_{k}^{1,0}\}$}\tabularnewline
\hline
\hline
\multirow{1}{*}{{\footnotesize $1$}} & {\footnotesize $(2l+2m+n+3)(2l-n)$;} & {\footnotesize $\{0,0,1\}$} & {\footnotesize $5+4l^{2}+2m+2l(2m+5)-2n(m+1)$} & {\footnotesize ${\color{black}\{0,2,1\}}$}\tabularnewline
\hline
{\footnotesize $2$} & \textbf{\footnotesize $n+m+2$} & {\footnotesize $\{1,0,1\}$} & {\footnotesize $4l(l+m)-2m(n+1)-n(n+5)-5$} & {\footnotesize $\{2,0,1\}$}\tabularnewline
\hline
{\footnotesize $3$} & {\footnotesize $-m(4l+m+1)$} & {\footnotesize $\{2,0,-1\}$} & {\footnotesize $-m(4l+m+1)$} & {\footnotesize $\{4,0,-1\}$}\tabularnewline
\hline
{\footnotesize $4$} & {\footnotesize $m(m+2n+1)$} & {\footnotesize $\{0,2,-1\}$} & {\footnotesize $m(m+2n+1)$} & {\footnotesize $\{0,4,-1\}$}\tabularnewline
\hline
\multirow{1}{*}{{\footnotesize $5$}} & {\footnotesize $n(n-1)$} & {\footnotesize $\{-2,2,1\}$} & {\footnotesize $n(n-1)$} & {\footnotesize $\{-2,4,1\}$}\tabularnewline
\hline
{\footnotesize $6$} & {\footnotesize $-2l(2l-1)$} & {\footnotesize $\{2,-2,1\}$} & {\footnotesize $-2l(2l-1)$} & {\footnotesize $\{4,-2,1\}$}\tabularnewline
\hline
{\footnotesize $7$} & {\footnotesize $-n$} & {\footnotesize $\{-1,2,1\}$} & {\footnotesize $-n$} & {\footnotesize $\{-1,4,1\}$}\tabularnewline
\hline
{\footnotesize $8$} & {\footnotesize $-m$} & {\footnotesize $\{1,2,-1\}$} & {\footnotesize $-m$} & {\footnotesize $\{1,4,-1\}$}\tabularnewline
\hline
{\footnotesize $9$} & {\footnotesize $1/4$} & {\footnotesize $\{0,2,1\}$} & {\footnotesize $1/4$} & {\footnotesize $\{0,4,1\}$}\tabularnewline
\hline
{\footnotesize $10$} & {\footnotesize $-1/4$} & {\footnotesize $\{2,0,1\}$} & {\footnotesize $-1/4$} & {\footnotesize $\{4,0,1\}$}\tabularnewline
\hline
{\footnotesize $11$} &  &  & {\footnotesize $2m(n-2l)$} & {\footnotesize $\{2,2,-1\}$}\tabularnewline
\hline
{\footnotesize $12$} &  &  & {\footnotesize $-m$} & {\footnotesize $\{3,2,-1\}$}\tabularnewline
\hline
{\footnotesize $13$} &  &  & {\footnotesize $n+m+3$} & {\footnotesize $\{3,0,1\}$}\tabularnewline
\hline
{\footnotesize $14$} &  &  & {\footnotesize $(m+1)$} & {\footnotesize $\{1,2,1\}$}\tabularnewline
\hline
\end{tabular}
\label{T1}
\end{center}
\end{table}

\newpage

\begin{thebibliography}{99}

\bibitem{HYL} E. A. Hylleraas, Z. Phys. \textbf{54}, 347-366 (1929).

\bibitem{B35} J. H. Bartlett, J. J. Gibbons and C. G. Dunn, "The Normal Helium Atom", Phys. Rev. \textbf{47}, 679-680 (1935).
\bibitem{B37} J. H. Bartlett, "The Helium Wave Equation", Phys. Rev. \textbf{51}, 661-669 (1937).

\bibitem{FOCK} V. A. Fock, Izv. Akad. Nauk SSSR, Ser. Fiz. \textbf{18}, 161 (1954).

\bibitem{HYM} E. A. Hylleraas and J. Midtdal, "Ground State Energy of Two-Electron Atoms", Phys. Rev. \textbf{103}, 829-830 (1956); "Ground State Energy of Two-Electron Atoms. Corrective Results", Phys. Rev. \textbf{109}, 1013-1014 (1958);

\bibitem{FRP} K. Frankowski and C. L. Pekeris, "Logarithmic Terms in the Wave Functions of the Ground State of Two-Electron Atoms", Phys. Rev. A \textbf{146}, 46-49 (1966);
    K. Frankowski, "Logarithmic Terms in the Wave Functions of the $2^1S$ and $2^3S$ State of Two-Electron Atoms", Phys. Rev. A \textbf{160}, 1-3 (1967).

\bibitem{FHM} D. E. Freund, B. D. Huxtable, and J. D. Morgan III, "Variatinal calculations on the helium isoelectronic sequence", Phys. Rev. A \textbf{29}, 980-982 (1984).

\bibitem{ES1} A. M. Ermolaev and G. B. Sochilin, "The Ground State of Two-Electron Atoms and Ions",
Sov. Phys. - Doklady \textbf{9}, 292-295 (1964); "$2^3S State of Helium$", Int. J. Quant. Chem. 2, 333-339 (1968).

\bibitem{GAM}
 P. C. Abbott and E. N. Maslen, "Coordinate systems and analytic expansions for three-body atomic wavefunctions: I. Partial summation for the Fock expansion in hyperspherical coordinates", J. Phys. A: Math. Gen. \textbf{20}, 2043-2075 (1987);
J. E. Gottschalk, P. C. Abbott and E. N. Maslen, "Coordinate systems and analytic expansions for three-body atomic wavefunctions: II. Closed form wavefunction to second order in $r$", J. Phys. A: Math. Gen. \textbf{20}, 2077-2104 (1987);
J. E. Gottschalk and E. N. Maslen, "Coordinate systems and analytic expansions for three-body atomic wavefunctions: III. Derivative continuity via solution to Laplace's equation", J. Phys. A: Math. Gen. \textbf{20}, 2781-2803 (1987).

\bibitem{FOR} R. C. Forrey, "Compact representation of helium wave functions in perimetric and hyperspherical coordinates", Phys. Rev. A \textbf{69}, 022504 (2004).

\bibitem{MYE} C. R. Myers, C. J. Umrigar, J. P. Sethna, J. D. Morgan III, "Fock's expansion, Kato's cusp conditions, and the exponential ansatz", Phys. Rev. A \textbf{44}, 5537-5546 (1991).

\bibitem{DRU} E. G. Drukarev, A. I. Mikhailov, I. A.  Mikhailov and W. Scheid, "Triple-coalescence singularity in a dynamical atomic process", JETP Lett. \textbf{86}, 702-704 (2007).

\bibitem{DRK} "Atomic, Molecular, \& Optical Physics Handbook", edited by G. W. F. Drake, AIP Press, New-York, 1996.

\bibitem{MAT}  http://www.wolfram.com/mathematica/

\bibitem{PRU} A. P. Prudnikov, Yu. A. Brychkov and O. I. Marichev, " Integrals and Series. Vol 1. Elementary Functions", Gordon and Breach S. P., New-York, 1986.

\bibitem{PRU3} A. P. Prudnikov, Yu. A. Brychkov and O. I. Marichev, " Integrals and Series. Vol 3. More Special Functions", Gordon and Breach S. P., New-York, 1986.

\bibitem{KK} E. D. Krupnikov, and K. S. K\"{o}lbig, "Some special cases of the generalized hypergeometric function $~_{q+1}F_q$", J. Comp. Appl. Math. \textbf{78}, 79-95 (1997).

\bibitem{LEZ2} E. Z. Liverts and N. Barnea , "S-states of helium-like ions",  Comp. Phys. Comm. \textbf{182}, 1790-1795 (2011).

\bibitem{LEZ3} E. Z. Liverts and N. Barnea, "Three-body systems with Coulomb interaction. Bound and
quasi-bound S-states", Comp. Phys. Comm. \textbf{184}, 2596-2603 (2013).

\end{thebibliography}
\end{document}